\documentclass[a4paper,11pt,twoside]{article}
\usepackage{fancyhdr,latexsym,amsmath,amsfonts,amssymb,amsbsy,amsthm,url}
\usepackage[hscale=0.7,centering]{geometry}
\usepackage{graphicx,epsfig}
\usepackage{amssymb,amsmath}
\usepackage{times}
\usepackage{color}
\usepackage{amsfonts}[11pt]
\usepackage{amsmath}[11pt]
\numberwithin{equation}{section}

\fancypagestyle{usual}{
  \lhead[\thepage]{}
  \chead[{\it \shortauthors}]{\sc \shorttitle}
  \rhead[]{\thepage}
  \lfoot[]{}
  \cfoot[]{}
  \rfoot[]{}
  
}

\makeatletter

\renewcommand{\section}{
  \@startsection
  {section}
  {1}
  {0pt}
  {1.1\baselineskip}
  {0.2\baselineskip}
  {\sc \centering}
}
\makeatother

\begin{document}
\setlength{\topmargin}{0.051in}
\newtheorem{lm}{Lemma}
\newtheorem{pr}{Proposition}
\title{\large\sc Analysis of Hepatitis C Viral Dynamics Using Latin Hypercube Sampling
}\normalsize
\author{\sc{Gaurav Pachpute} \thanks{Department of Mathematics, Indian Institute of Technology Guwahati, Guwahati 781039, Assam, India,
e-mail: g.pachpute@iitg.ernet.in} 
\and \sc{Siddhartha P. Chakrabarty} \thanks{Department of Mathematics, Indian Institute of Technology Guwahati, Guwahati 781039, Assam, India,
e-mail: pratim@iitg.ernet.in}}
\date{}
\maketitle
\begin{abstract}
We consider a mathematical model comprising of four coupled ordinary differential equations (ODEs) for studying 
the hepatitis C (HCV) viral dynamics. The model embodies the efficacies of a combination therapy of interferon and ribavirin.
A condition for the stability of the uninfected and the infected steady states is presented. A large number of sample points
for the model parameters (which were physiologically feasible) were generated using Latin hypercube sampling. Analysis of our
simulated values indicated approximately $24 \%$ cases as having an uninfected steady state. Statistical tests like the $\chi^{2}$-test 
and the Spearman's test were also done on the sample values. 
The results of these tests indicate a distinctly differently distribution of certain parameter values and not in case of others, 
vis-a-vis, the stability of the uninfected and the infected steady states.\\


\end{abstract}

\section{Introduction}

Hepatitis C, which is an infectious disease caused by the hepatitis C virus (HCV), has a widespread global prevalence.
It is estimated to have infected 170 million individuals worldwide \cite{Razali07,Dixit08}.
HCV is usually transmitted through blood contact with an infected person. In places like United States
and Europe, the primary mode of transmission of HCV is through injected drug usage \cite{Razali07}.
In India however, the lack of effective and reliable anti-HCV screening amongst blood donors 
is a major source of infection \cite{Pal02,Mukhopadhya08}. New HCV infections (through intravenous drug 
usage or otherwise) can be primarily classified as acute and chronic \cite{Razali07}. While the acute cases
in general do not have major detrimental implications and the virus is cleared, the ramifications of chronic cases
are far reaching \cite{Razali07}. It could lead to the occurrence of liver cirrhosis and may eventually lead to
hepatocellular carcinoma (HCC). About $50-80\%$ of HCV infections are chronic in nature \cite{Roe08}. Of these $10-20\%$
develop liver cirrhosis of which about $5 \%$ are likely to be afflicted with HCC \cite{Roe08}.

The current treatment for HCV infection involves the combination therapy of pegylated interferon (IFN) and ribavirin
\cite{Dixit08}, which yields long term response in only $50 \%$ of the cases. There is very little therapeuitic 
alternative for the cases of non-responders. Feld et al. \cite{Feld05} dwell on the mechanism of action of 
combination treatment of IFN and ribavirin in HCV infected patients.
They report limited success ($6-12 \%$ for a six-month treatment and $16-20\%$ for a one-year treatment) in case of
an IFN based monotreatment. The combination therapy however resulted in significant improvement
(more than $50 \%$) in response rates. Perelson et al. \cite{Perelson05} in their review article, discuss the role
of this combination therapy for HCV infection, taking into account the fall in the efficacy levels of the drugs
between dosing intervals.

Several mathematical models have been proposed for the study of hepatitis C viral dynamics.
One of the earliest mathematical models was proposed by Neumann et al. \cite{Neumann98}, wherein
the dynamics of HCV and the effect of interferon-$\alpha$-2b were studied in vivo.
The quantitative representation of the dynamics involved the incorporation of some aspects of the earlier 
successful models for HIV and HBV \cite{Dixit08,Neumann98}. The model involved three coupled ODEs, where the 
key factors were identified as uninfected hepatocytes, productively infected hepatocytes and free HCV virions.
The model assumed the growth of the uninfected hepatocytes at a constant rate accompanied by a natural death rate.
In addition, the HCV was modeled as infecting the hepatocytes, 
which also had a natural death rate. The infected hepatocytes in turn abetted the growth of HCV.
In addition, a term for the efficacy of IFN was incorporated to reduce the rate of infection of hepatocytes and
to block the production of HCV from infected hepatocytes.
The model exhibited a decline in the levels of infected hepatocytes and HCV accompanied by an increase in the level 
of uninfected hepatocytes. A single phasic decline was observed in case the efficacy of IFN in blocking
the production of virions  was taken to be zero, which is inconsistent with clinical observations \cite{Dixit08,Neumann98}. 
Otherwise it showed a biphasic decline in viral load which is more realistic from the biomedical point of view.
The model indicated that the key role of IFN is in blocking the production of virions from infected cells as compared to blocking 
infection of hepatocytes, which is minimal \cite{Dixit08,Neumann98}.

Dixit et al. \cite{Dixit04} extended the work of Neumann et al. \cite{Neumann98} in the sense of
explicitly including the action of ribavirin. In this model \cite{Dixit08,Dixit04} the virion
population is divided into infectious and non-infectious. Dixit et al. \cite{Dixit04}
assume that ribavirin (either on its own or in conjunction with IFN)
renders a fraction of the newly produced HCV non-infectious. Another assumption of the model
(in contrast to Neumann et al. \cite{Neumann98}) was that the role of IFN in blocking the 
production of infected hepatocytes is not significant. The model predicts that ribavirin does not have 
an impact on the IFN induced first phase decline \cite{Dixit08}.
If the efficacy of IFN is large enough, then the second phase decline is not significantly affected by ribavirin. 
However, if the IFN efficacy is much smaller than $1$, then the impact of ribavirin in the
second phase decline is more profound. A plausible explanation for this is that a high
IFN efficacy results in low HCV levels, which in turn reduces the role that ribavirin plays in rendering
the virions as non-infectious. The model was able to explain why ribavirin enhances the second phase of decline
in some cases and not in others \cite{Dixit08}.

Dahari et al. \cite{Dahari07} further advanced the work of Neumann et al. \cite{Neumann98} and Dixit et al. \cite{Dixit04}.
Their model \cite{Dahari07} incorporated the proliferation of both infected and uninfected hepatocytes which was not considered
in the earlier models. They include density dependent proliferation for both infected and uninfected hepatocytes, which
restricted the growth of liver to a maximum possible size. This model was able to explain the limitations of 
the model of Dixit et al. \cite{Dixit04}, which could not explain the non-response of some patients and
the triphasic decline patterns. The model defined a critical threshold efficacy below which
there cannot be sustained long term viral load decline \cite{Dixit04,Dahari07}.
In then case when the efficacy is above the threshold one observes a decline in the viral load, which could be
biphasic or triphasic. The triphasic decay can be attributed to the homeostatic liver regeneration.

\section{Mathematical Model}

The model under consideration is a fusion of the earlier models \cite{Neumann98,Dixit04,Dahari07,Dahari07b}
and is given by the following system of four coupled ODEs,
\begin{eqnarray}
\label{modeleq}
\frac{dT}{dt}&=&s+rT\left(1-\frac{T+I}{T_{\max}}\right)-dT-\beta T V_{I} \nonumber\\
\frac{dI}{dt}&=&\beta TV_{I}+rI\left(1-\frac{T+I}{T_{\max}}\right)-\delta I \nonumber\\
\frac{dV_{I}}{dt}&=&(1-\rho)(1-\epsilon_{p})p I-cV_{I}\nonumber\\
\frac{dV_{NI}}{dt}&=&\rho(1-\epsilon_{p})pI-cV_{NI}
\end{eqnarray}
Here $T(t)$ and $I(t)$ are the number of uninfected and infected hepatocytes respectively while
$V_{I}(t)$ and $V_{NI}(t)$ are the number of infectious and non-infectious virions (corresponding to HCV RNA genome equivalents)
respectively. The assumption in the model is that the hepatocytes are being produced at a constant rate $s$
and have a natural death rate of $d$, while proliferating at a rate $r$ with $T_{\max}$
being the maximum possible hepatocyte (both uninfected and infected) population level. The hepatocytes are assumed to be infected 
by the virions at a rate $\beta$. The proliferation of infected hepatocytes is also assumed to take place at the rate $r$
with a natural death rate $\delta$.
In the absence of any kind of treatment, the infected hepatocytes produce infectious virions at a rate $p$.
The administration of IFN lowers this production by a factor of $(1-\epsilon_{p})$, where $\epsilon_{p}$ is the
efficacy of IFN. Finally, the model incorporates the efficacy $\rho$ of ribavirin which renders a fraction of the infectious virions
non-infectious, with a death rate $c$ for both these populations.
The model under consideration admits two steady states, {\it viz.} the uninfected and the infected steady states, 
as given below \cite{Dahari07},
\begin{enumerate}
\item \[T^{(u)}= \frac{T_{\max}}{2r}\left[r-d+\sqrt{(r-d)^{2}+\frac{4rs}{T_{\max}}}\right],
I^{(u)}=0, V_{I}^{(u)}=0,V_{NI}^{(u)}=0.\]
\item 
\begin{eqnarray*}
T^{(i)}&=&\frac{1}{2}\left[-D+\sqrt{D^{2}+\frac{4sT_{\max}}{rA^{2}}}\right]\\
I^{(i)}&=&T^{(i)}(A-1)+T_{\max}-B\\
V_{I}^{(i)}&=&\frac{(1-\epsilon_{p})(1-\rho)pI^{(i)}}{c},V_{NI}^{(i)}=\frac{\rho V_{I}^{(i)}}{(1-\rho)},
\end{eqnarray*}
where,
\[A=\frac{(1-\epsilon_{p})(1-\rho)p\beta T_{\max}}{cr}, B=\frac{\delta T_{\max}}{r}, 
D=\frac{1}{A}\left[T_{\max}+\frac{dB}{\delta A}-B\left(\frac{1}{A} + 1\right)\right]\].
\end{enumerate}
Under the physiological conditions of $r>d$ and $s\le dT_{\max}$ it can be shown \cite{Dahari07}
that the condition for stability of the uninfected steady state is,
\[(1-\epsilon_{p})(1-\rho)<\frac{c(\delta T_{\max}+rT^{(u)}-rT_{\max})}{p\beta T^{(u)}T_{\max}},\]
while that of the infected steady state is,
\[(1-\epsilon_{p})(1-\rho)>\frac{c(\delta T_{\max}+rT^{(u)}-rT_{\max})}{p\beta T^{(u)}T_{\max}}.\]
Thus there is a transcritical bifurcation at,
\[(1-\epsilon_{p})(1-\rho)=\frac{c(\delta T_{\max}+rT^{(u)}-rT_{\max})}{p\beta T^{(u)}T_{\max}}.\]

\section{Latin Hypercube Sampling}

\def\muvector{\stackrel{\rightarrow}{\mu}}

In this article we use the method of Latin hypercube sampling in order to generate a collection of parameter values
from a multivariate normal distribution with a specified mean vector $\muvector$ and covariance matrix $\Sigma$.
The theoretical and computational aspects of this method can be found in the pioneering work of McKay et al. \cite{McKay79} 
which was further advanced by Stein \cite{Stein87}. Latin hypercube sampling is a multidimensional version of unidimensional stratification.
\cite{Glasserman04}. Latin hypercube sampling is used to generate parameter values from a given distribution. In this method, 
random variates are generated from a $d$-dimensional uniform distribution over the hypercube $[0,1)^{d}$. Random variates from
other $d$- dimensional distributions can be generated from the uniform variates using the inverse transform method \cite{Glasserman04}.
Stratification in one-dimensional uniform distribution can be achieved by dividing the interval into $K$ strata. The extension of this
idea to multidimensional stratification is not trivial as the sample size of $K^{d}$ is computationally expensive even for moderate values of $d$
(since $K$ has to be large enough). We present a brief outline of this method (after Glasserman \cite{Glasserman04}).

To begin with, we generate random variates $V_{i}^{(j)}$ from a uniform distribution over $[(j-1)/K,j/K)$ for $i=1,2,\dots,d$ 
and a set of independent permutations $\sigma_{1},\dots,\sigma_{d}$ over the set $\{1,2,\dots,K\}$(from $K !$ possibilities). 
The vectors $\left(V_{1}^{(j)},V_{2}^{(j)},\dots,V_{d}^{(j)}\right)^{\top}$ (for $j=1,2,\dots,K$) represent uniformly distributed 
points over the hypercube $[0,1)^{d}$. We can then set $V_{i}^{j}\leftarrow V_{i}^{\sigma_{i}(j)}$ which gives us a stratified sample 
over the hypercube. An obvious way to achieve this is to set, 
\[V_{i}^{(j)}=\frac{\sigma_{i}(j)-1+U_{i}^{(j)}}{K}, \: i=1,\dots,d, \: j=1,\dots,K, \: \text{and} \: 
U_{i}^{(j)}\sim \text{Uniform}\: [0,1)\]
In generating this construction, one of the crucial assumptions is the independence of marginals, for introducing a correlation 
between parameter distributions might alter the stratification properties of Latin hypercube sampling. This is very much evident in generating a 
sample from a normal distribution which does not have a diagonal covariance matrix. The values so generated will not generally be stratified.
In order to generate from (say) a normal distribution with mean vector $\muvector=(\mu_{i})_{i=1}^{d}$ 
and diagonal covariance matrix $\Sigma=(\Sigma_{ij})_{i,j=1}^{d}$ we set,
\[Z_{i}^{(j)}=\Phi_{\mu_{i},\Sigma_{ii}}^{-1}\left(V_{i}^{(j)}\right), i=1,\dots,d, j=1,\dots,K\] 
where $\Phi_{\mu_{i},\Sigma_{ii}}$ is the one-dimensional cumulative normal distribution with mean $\mu_{i}$ and variance $\Sigma_{ii}$.

\section{Results and Discussion}

\def\SS{\scriptsize}
\def\TM{\!\!$^{\mbox{\texttt{\SS TM}}}$~}

We implemented the Latin hypercube sampling procedure describe above in MatLab \TM and generated $K=2^{15}$ sample points
for all the parameter values. The mean vector $\muvector$ was taken to be the values given in Table \ref{tableone}, \textit{i.e,}
$\mu=\left(s,d,p,\beta,c,\delta,r,T_{\max}\right)^{\top}$. The diagonal covariance matrix $\Sigma$ was taken as 
$\text{diag}\left(s^2,d^2,p^2,\beta^2,c^2,\delta^2,r^2,T_{\max}^{2}\right)$.
Once the $K=2^{15}$ sets of parameter values were generated, they were tested for positivity as well as
the physiological conditions ($r>d$ and $s\le d T_{\max}$) as given in Section 2. This test left us with $8213$ sample parameter sets
that were feasible.

A number of tests were performed to understand how the sample parameter values correlate. Firstly, a $\chi^{2}$ test was 
performed on the $8213$ accepted sample points against the frequency of positive samples from their 
respective intended distributions under the \emph{goodness of fit} null hypothesis. The null hypothesis (\textit{set of sample 
points which are biomedically feasible do not deviate from the intended distribution}), was not rejected for any of the sample parameters. 
On the other hand, a test done under the \emph{independence} null hypothesis rejects this hypothesis for all parameters with respect to the 
significance level chosen (P $< 0.05$). Since, physiological conditions add "restrictions" to the intended distribution, we can surmise 
from these tests that the accepted values do not deviate from this distribution.

The accepted sample points were categorized into three parts depending on whether the value
\[Cr^{*}=\frac{c(\delta T_{\max}+rT^{(u)}-rT_{\max})}{p\beta T^{(u)}T_{\max}},\]
was greater than $1$, less than $0$ or was in the interval $[0,1]$.
Notice that, when $0 \le Cr^{*} \le 1$, both steady states can be stable on two disjoint sets of feasible 
$\epsilon_{p}$ and $\rho$ values (due to the transcritical bifurcation).
The number of such sample points, which emerged from our simulations, was $6262~(76.25\%)$.
On the other hand, when $ Cr^{*}> 1$ or $Cr^{*} < 0$, only one of the two steady states is stable. 
The sample parameter values, which gave the $Cr^{*}$ values to be greater than $1$ can only lead to the stability of
the uninfected steady state, since in this case the stability condition for the uninfected steady state is satisfied for all
possible values of $\epsilon_{p}$ or $\rho$.  
The percentage of cases with $Cr^{*}>1$ was $23.41\%$ ($1923$ out of $8213$), which is reasonly
in line with biomedical observations \cite{Roe08,Dahari07}. 
Similarly, the cases where $Cr^{*}<0$ were considered to the ones 
for which sustained virological response cannot be achieved. 
Samples from each of these sets were compared against the frequencies of the biomedically accepted samples.

\begin{figure}[hb]
\centering
\includegraphics[width=0.7\textwidth]{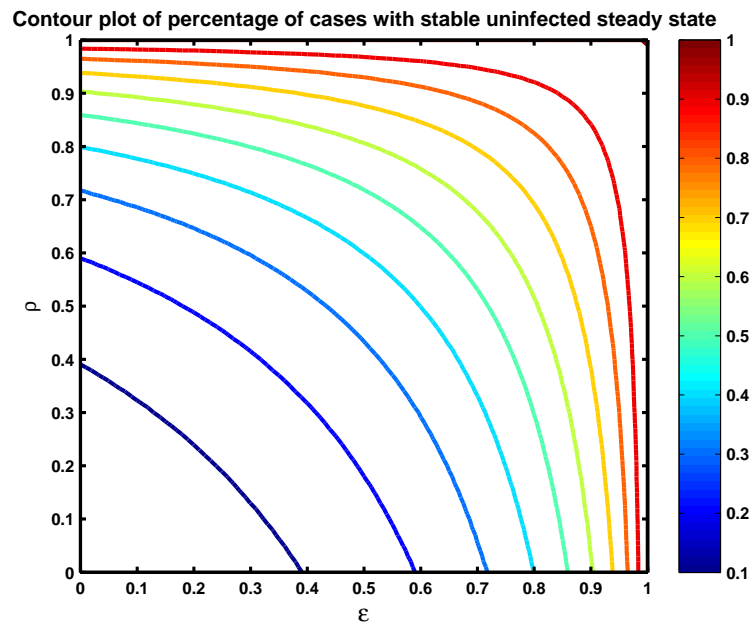}
\caption{Contour plot of percentage of cases with stable uninfected steady state}
\label{contourplot}
\end{figure}

\begin{figure}[hb]
\centering
\includegraphics[width=0.7\textwidth]{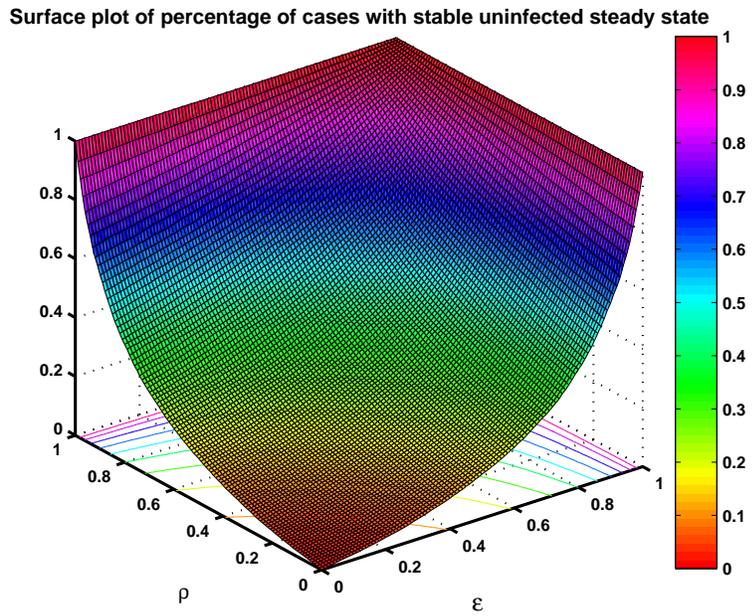}
\caption{Surface plot of percentage of cases with stable uninfected steady state}
\label{surfaceplot}
\end{figure}

\begin{figure}[hb]
\centering
\includegraphics[width=0.7\textwidth]{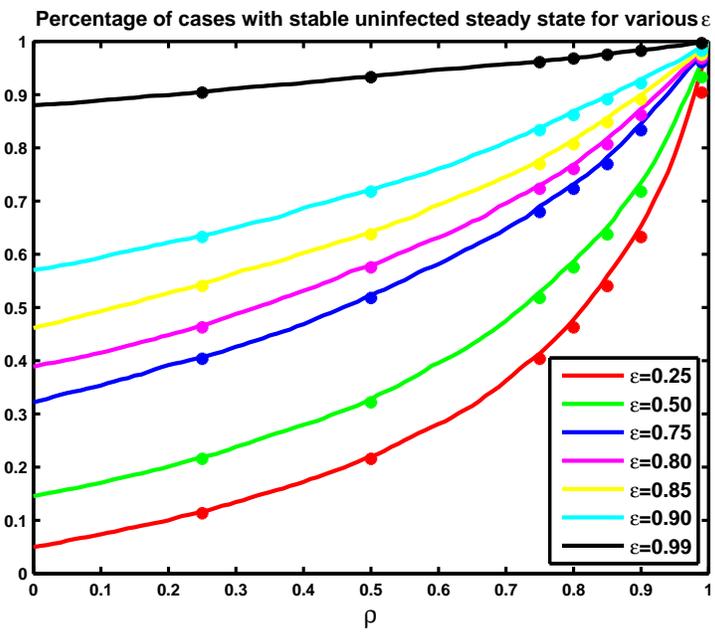}
\caption{Percentage of cases with stable uninfected steady state for various $\epsilon$}
\label{rho}
\end{figure}

\begin{figure}[hb]
\centering
\includegraphics[width=0.7\textwidth]{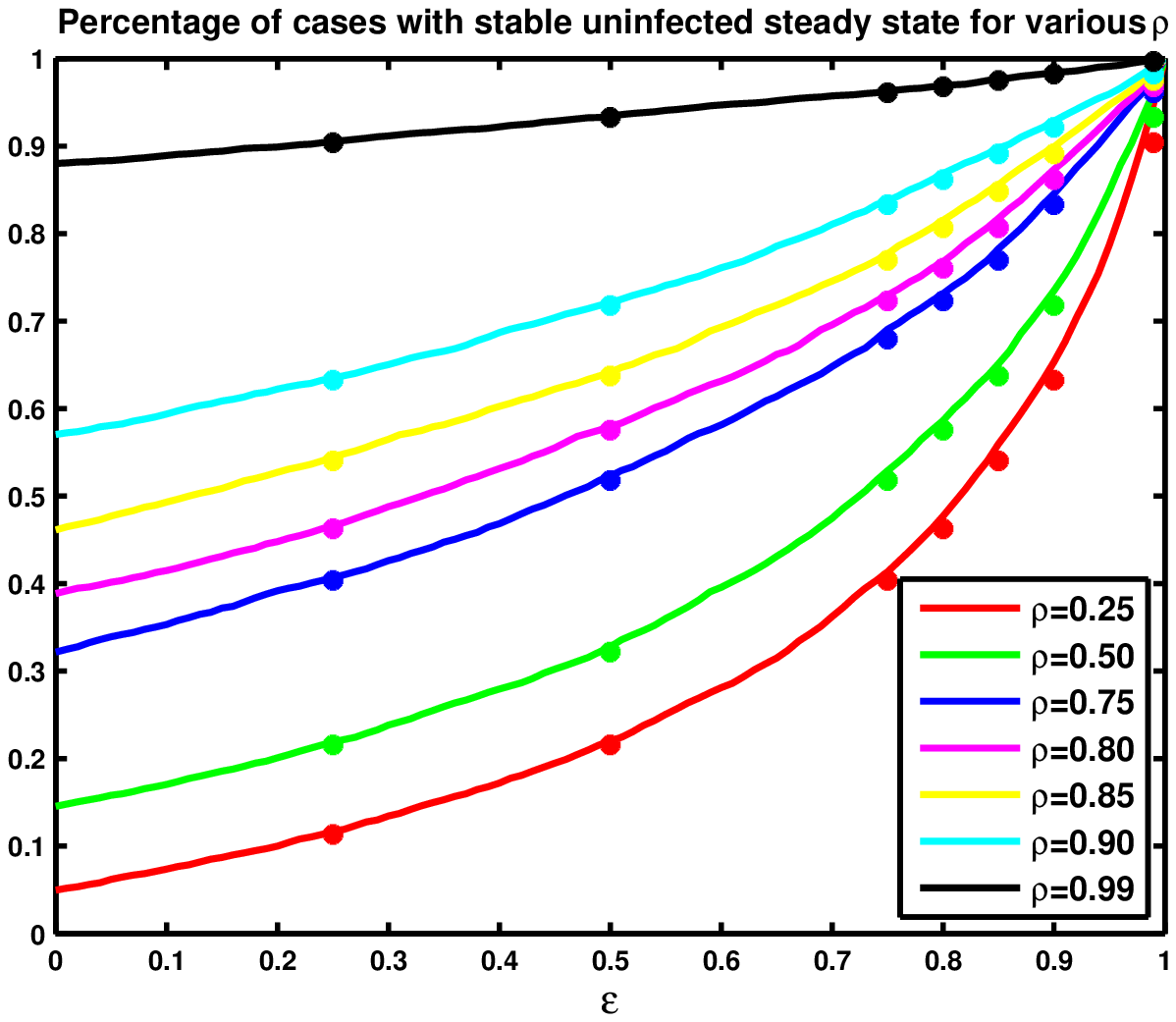}
\caption{Percentage of cases with stable uninfected steady state for various $\rho$}
\label{epsilon}
\end{figure}

We considered values of $\epsilon_{p}$ and $\rho$ in the range $[0,1]$ and in increments of $0.01$.
For these pair of $101\times 101$ such values we kept a count of the number of values of sample points which satisfied the
condition $(1-\epsilon_{p})(1-\rho)<Cr^{*}$. Recall that, this condition corresponds to the stability of the uninfected steady state.
Thus, this gives us the cumulative distribution of the percentage of cases which give a stable uninfected steady state for the
corresponding drug efficacies. The results for these are presented in a contour plot in Figure (\ref{contourplot})
and in a surface plot  in Figure (\ref{surfaceplot}).
We plotted the results for the various values of $\epsilon_{p}$ against $\rho$ (Figure \ref{rho}) and
for values of $\rho$ against $\epsilon_{p}$ (Figure \ref{epsilon}). One observes 
that for smaller values of IFN efficacy $\epsilon_{p}$, the response (in terms of percentage of cases with stable
uninfected steady state) is very sensitive to change in the efficacy $\rho$ of ribavirin. The impact of $\rho$ however is much
less evident in case of IFN being highly effective. This observation is consistent with the findings of Dixit et al. \cite{Dixit04}.

{\SS
\begin{center}
\begin{table}[h!]
\begin{center}
\begin{tabular}{|c|c|}
\hline
Parameter& Value \cite{Dahari07}\\
\hline
$s$ & $1.0$ cell ml$^{-1}$ day$^{-1}$, \\
$d$ & $0.01$ day$^{-1}$,\\
$p$ & $2.9$ virions day$^{-1}$, \\
$\beta$ & $2.25\times 10^{-7}$ ml day$^{-1}$ virions$^{-1}$, \\
$c$ & $6.0$ day$^{-1}$\\
$\delta$ & $1.0$ day$^{-1}$\\
$r$ & $2.0$ day$^{-1}$\\
$T_{\max}$ & $3.6\times 10^{7}$ cells ml$^{-1}$,\\
\hline
\end{tabular}
\caption{Mean parameter values for the Latin hypercube sampling
\label{tableone}}
\end{center}
\end{table}
\end{center}
}

For the samples which have $Cr^{*}$ values greater than $1$ or between $0$ and $1$, we observe that for variables 
$p$, $\beta$, $c$, $\delta$ and $T_{\max}$, the null hypothesis (\textit{independence}) is rejected. 
It implies that the sample points from these two sets tend to deviate more from their original distribution
which leads them to be more restricted in nature.
Sample points which have a negative $Cr^{*}$ value predominantly depend upon the death rates $d$ and $\delta$ of the 
uninfected and infected hepatocytes respectively, for which the null hypothesis fails.
Clearly, these are cases exhibiting larger values of $d$ and smaller values of $\delta$. 
In biological terms, these cases present with a higher clearance rate of uninfected hepatocytes and a 
smaller rate of infected hepatocytes. It can also be inferred from equation (\ref{modeleq}) that there is little role for drug efficacies 
if the above conditions are prevalent. Also, $\delta$ does not follow the original distribution in any of the cases, 
indicating its importance in the distinction 
amongst the three cases.

Next, we chose a threshold value $0\le Cr_{thr}\le 1$ and analyzed the independence of different parameters which have $Cr^{*}$ values
on both sides of $Cr_{thr}$. 
Notice that if $Cr^{*}>Cr_{thr}$, the uninfected steady state can be reached for smaller values of drug efficacies as compared to
$Cr^{*}<Cr_{thr}$. By varying the value of $Cr_{thr}$ from 0.2 to 0.4, a Spearman's test showed a modest monotonic relationship 
between $p$ and $\beta$ for the samples with $Cr^{*}>Cr_{thr}$ and not in case of $Cr^{*}<Cr_{thr}$. 
Biologically, $\beta$ represents the rate of infection and $p$ represents the 
rate with which infected hepatocytes are converted into virions. The results highlight the fact that for the uninfected steady state to 
be easily reachable, these two rates cannot be simultaneously high. 

\section{Conclusion}

In Latin hypercube sampling, one is at the liberty to choose distributions which best fits the needs (such as previous findings, 
physiological constraints \textit{etc.}). In this paper, we applied the Latin hypercube sampling to generate a large number 
of sample parameter values for a HCV model. Once the sampling was done, we chose values that were positive and satisfied the 
physiological conditions. One may perform \textit{goodness of fit} and \textit{independence} tests at every stage of adding constraints to 
test how much and in what way they affect the distribution of the generated samples.
This lets us better understand the implications of said constraints on a large population of sample parameters.
In our case, the values found feasible were subject to the $\chi^{2}$-test and the Spearman's test.
The $\chi^{2}$ tests, were very useful in checking the \textit{goodness of fit and independence} of a parameter against 
an intended distribution. On the other hand, Spearman's test was used to check for a monotonic relationship between parameters.
The methods and tests mentioned in this paper which were also performed on a sample generated from the log-normal distribution 
gave similar results.

\DeclareGraphicsExtensions{.eps}

\begin{thebibliography}{99}

\bibitem{Razali07}
Razali K. et al. (2007) Modelling the hepatitis C virus epidemic in Australia.
Drug and Alcohol Dependence 91: 228-235.

\bibitem {Dixit08} Dixit NM (2008) Advances in the mathematical modelling of hepatitis C virus dynamics.
Journal of the Indian Institute of Science 88(1):37–43.

\bibitem{Pal02} Pal SK, Chalamalasetty BK, Choudhuri G (2002) Hepatitis C: a major health problem of India.
Current Science 83(9): 1058-1059.

\bibitem{Mukhopadhya08} Mukhopadhya A (2008) Hepatitis C in India.
Journal of Biosciences 33(4): 465-473.

\bibitem{Roe08} Roe B, Hall WW (2008) Cellular and molecular interactions in coinfection with hepatitis C virus and human immunodeficiency virus.
Expert Reviews in Molecular Medicine 10: null-null.

\bibitem{Feld05} Feld JJ, Hoofnagle JH (2005) Mechanism of action of interferon and ribavirin in treatment of hepatitis C.
Nature 436:967–972.

\bibitem{Perelson05} Perelson AS, Herrmann E, Micol F, Zeuzem S (2005), New kinetic models for the hepatitis C virus, Hepatology 42(4):749–754.

\bibitem{Neumann98} Neumann AU, Lam NP, Dahari H, Gretch DR, Wiley TE, Layden TJ, Perelson AS. (1998)
Hepatitis C viral dynamics in vivo and the antiviral efficacy of interferon-$\alpha$ therapy. Science 282: 103-107.

\bibitem{Dixit04} Dixit NM, Layden-Almer JE, Layden TJ, Perelson AS (2004) Modelling how ribavirin improves interferon response rates
in hepatitis C virus infection. Nature 432:922–924.

\bibitem{Dahari07} Dahari H, Lo A, Ribeiro RM, Perelson AS (2007) Modeling hepatitis C virus dynamics: liver regeneration and critical drug efficacy.
Journal of Theoretical Biology 247: 371-381.

\bibitem{Dahari07b} Dahari H, Ribeiro RM, Perelson AS (2007), Triphasic decline of hepatitis C virus RNA during
antiviral therapy, Hepatology, 46(1): 16-21.

\bibitem{McKay79} McKay MD, Beckman RJ, Conover, WJ (1979), 
A comparison of three methods for selecting values of input variables in the analysis of output from a computer code,
Technometrics, 21(2): 239-245.

\bibitem{Stein87} Stein M (1987), Large sample properties of simulations using latin hypercube sampling, Technometrics, 29(2): 143-151.

\bibitem{Glasserman04} Glasserman P (2004), Monte Carlo methods in financial engineering, Springer.

%
%
%
%
%
%
%
%
%
%
%
%
%
%
%
%
%

\end{thebibliography}
\end{document}